\documentclass{article}
\usepackage{spconf,amsmath,graphicx,hyperref}
\usepackage{xcolor}
\usepackage{amssymb}
\usepackage{graphicx}
\usepackage{subcaption}
\usepackage{cite}

\def\b1{{\boldsymbol{1}}}
\def\c1{{\textcircled{a}}}

\def\bn{{\boldsymbol{n}}}

\def\bx{{\boldsymbol{x}}}
\def\by{{\mathbf{y}}}

\def\bA{{\mathbf{A}}}

\def\bC{{\boldsymbol{C}}}

\def\bF{{\mathbf{F}}}

\def\bI{{\mathbf{I}}}

\def\bS{{\boldsymbol{S}}}

\title{CASCADED POSTERIOR SAMPLING WITH DIFFUSION PRIORS FOR HIGHLY
ACCELERATED MRI}
\name{Jyothi Rikhab Chand, Sahil Murtaza, Patrick Nicholas Nowak, Mathews Jacob}
\address{Department of Electrical and Computer Engineering, University of Virginia, VA, USA}
\begin{document}
%
\maketitle
\begin{abstract}
Diffusion models are powerful priors for inverse problems, including accelerated MRI. However, modeling the distribution of high-resolution images requires large networks with high computational cost. We propose a posterior sampling framework that uses cascaded diffusion models as multiscale priors. Conditioned on a coarse image, the distribution of fine-scale detail is local and well conditioned, and it can therefore be captured by a small network. Similarly, a small unconditional network is sufficient to represent low-resolution images, and it is used to recover the coarse image from the k-space center. This estimate then conditions the high-resolution prior, which is combined with all measurements to recover the final image. Both priors are trained on images alone, so they apply to any sampling pattern, and data consistency remains linear at every scale. In multi-coil brain MRI, the proposed method improves on the best single-scale diffusion baseline by 4–6 dB in PSNR at 15× acceleration, while each of its separately trained networks is about 10× smaller. The gain grows as the problem becomes more ill-posed, and ablations show that it arises from informative low-resolution conditioning.
\end{abstract}
\section{Introduction}
\label{sec:intro}


Diffusion models learn complex high-dimensional image distributions \cite{song2021score,karras2022edm} and are now widely used as priors for inverse problems in imaging \cite{daras2024survey}. Because a trained diffusion model is not tied to a specific acquisition operator, the same model can be combined with different forward models in a plug-and-play (PnP) manner \cite{dps,daps,alps}. This flexibility is a key advantage over end-to-end (E2E) methods \cite{sriram2020end,aggarwal2019modl,deepen}, which are trained for a particular acquisition setting.

Most diffusion priors rely on a single network that operates on the full-resolution image, which is challenging for large images. The coarse image structure is formed at high noise levels, where the denoiser must capture long-range dependencies across the entire image. At full resolution, this requires large networks with global receptive fields \cite{kadkhodaie2023learning}, and hence high computational cost. Posterior sampling \cite{daras2024survey} adds a further difficulty. The guidance from the measurements is least accurate at high noise levels, which is where the coarse image structure is determined. Consequently, at high acceleration, the coarse anatomy may be recovered incorrectly, even when the the k-space center is densely sampled. Many inverse problems hence rely on latent diffusion models, which reduce the computational cost by operating in the lower-dimensional latent space of an autoencoder \cite{rombach2022high}. However, the reconstruction accuracy is then bounded by the autoencoder, and the decoder makes the data-consistency update nonlinear \cite{song2024solving}.

In this work, we rely on multiscale diffusion models, which can overcome these challenges while preserving pixel-space data consistency. These models exploit the redundancy of images across scales, which had been widely used in image processing for image compression \cite{shapiro1993embedded,buccigrossi1999image}. Recent generative models extend this idea through coarse-to-fine factorizations \cite{ho2022cascaded,tian2024visual}, in which a fine-scale diffusion model is conditioned a coarse-scale image. Given the coarse image, the distribution of fine-scale details is better conditioned and more local than that of the full image, and hence it can be captured by small networks with limited receptive fields. Only the coarse-scale model must capture the long-range structure that is formed at high noise levels. Because the coarse image is low-dimensional, this global model can also be small \cite{guth2022wavelet,kadkhodaie2023learning}. This factorization is particularly well suited to accelerated MRI, where sampling patterns acquire the center of k-space densely and the periphery sparsely \cite{lustig2007sparse}. A low-resolution image can therefore be recovered from the central samples as a separate, much better-posed inverse problem, rather than through the high-noise steps of a full-resolution sampler. This coarse estimate provides reliable, measurement-derived information that constrains the fine-scale prior, where most of the ambiguity lies. 


In this paper, we introduce cascaded posterior sampling, which uses cascaded diffusion models \cite{ho2022cascaded} as measurement-informed multiscale priors. In the first stage, a low-resolution diffusion prior recovers a coarse image from the central k-space samples. In the second stage, this coarse estimate conditions a high-resolution diffusion prior, which is combined with all measurements to recover the final image. Each prior is about 
10× smaller than a single-scale model. Because the priors are trained and used in inference separately, the peak memory and computational complexity are governed by one small network. In multi-coil MRI, the proposed method matches single-scale diffusion baselines at moderate acceleration and improves on the best baseline by 4--6~dB in PSNR at 
15× acceleration. Denoising experiments confirm that the gain comes from informative low-resolution conditioning rather than from the additional network input.

\section{Background}

The proposed framework applies broadly to computational imaging inverse problems; in this work, we focus on accelerated MRI. Let $\bx \in \mathbb{C}^{m}$ denote the unknown high-resolution
MR image to be recovered from the acquired undersampled multi-coil k-space measurements $\by\in \mathbb{C}^{n}$. The measurement model is given by:
\begin{equation}
    \by = \bA \bx + \bn
\end{equation}

where $\bA = \bS\bF\bC$ denotes the multi-coil forward operator, $\bS$ is the sampling mask, $\bF$ is the Fourier matrix, $\bC$ is the coil sensitivity map, and $\bn \sim \mathcal{N}(0, \eta^{2}\bI)$. Under a Bayesian formulation, one can recover an MR image by sampling from the posterior distribution $p(\bx|\by)\propto p(\by|\bx) p(\bx)$, where $p(\by|\bx)$ denotes the likelihood and $p(\bx)$ denotes the MR image data distribution. Recently, diffusion models have emerged as powerful priors and have been adapted to sample from the posterior distribution. Given a diffusion model trained on MR images, the score network provides an estimate of the score of the perturbed image distribution. Several diffusion-based posterior sampling algorithms have been proposed for solving inverse problems by combining the learned diffusion model with the measurement likelihood \cite{dps,song2024solving,daps,alps}. For more details, see \cite{daras2024survey}.

\section{Cascaded posterior sampling}
Rather than modeling the full-resolution image distribution directly, we use a hierarchical prior that decomposes it into two simpler priors \cite{ho2022cascaded}:
\begin{equation}
    p(\bx)=\int p_{\phi}(\bx \mid \bx_{\mathrm{LR}})\, p_{\theta}(\bx_{\mathrm{LR}})\,d\bx_{\mathrm{LR}},
    \label{eq:hierarchical_prior}
\end{equation}
where $\bx_{\mathrm{LR}}$ denotes a low-resolution representation of $\bx$, $p_{\theta}(\bx_{\mathrm{LR}})$ denotes the low-resolution prior, and $p_{\phi}(\bx \mid \bx_{\mathrm{LR}})$ denotes the conditional high-resolution prior. Multiscale generative models sample from priors of this form by ancestral sampling: a coarse image is first drawn from the low-resolution prior, and the fine-scale details are then drawn conditioned on it \cite{ho2022cascaded,guth2022wavelet,kadkhodaie2023learning}. The factorization also simplifies learning in two ways. First, $p_{\theta}$ models images with $16\times$ fewer pixels, so its score is easier to learn and to sample. Second, the conditional distribution $p_{\phi}(\bx\mid\bx_{\mathrm{LR}})$ is better conditioned and more local than $p(\bx)$, because the coarse image already explains the long-range structure \cite{guth2022wavelet,kadkhodaie2023learning}. As a result, the fine scale can be modeled by a small network with a limited receptive field. Each of our priors has about 6.7M parameters and no attention layers, which reduces training and inference cost. In addition, both priors are trained on images alone. The acquisition operator enters only through the likelihood, and hence the same pair of priors applies to any sampling pattern.

\begin{figure}[t]
\centering
    \includegraphics[trim={0cm 0cm 0cm 0cm},clip,width=0.5\textwidth]{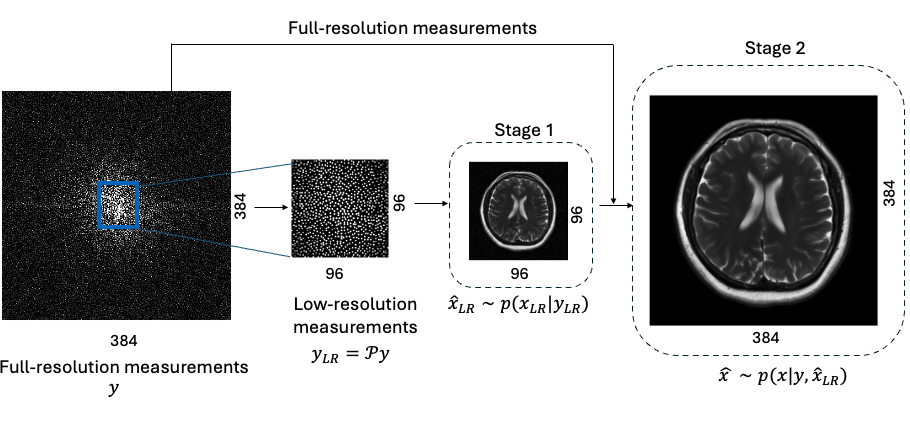}
    \caption{\small{Overview of the proposed two-stage cascaded diffusion reconstruction framework. Stage 1 reconstructs $\hat{\bx}_{\mathrm{LR}}$ from the central k-space samples $\by_{\mathrm{LR}}=\mathcal{P}\by$. Stage 2 uses $\hat{\bx}_{\mathrm{LR}}$ as a conditioning input, together with the full-resolution measurements $\by$ to reconstruct $\hat{\bx}$.}}
    \label{posterior_learning}
\end{figure}

Combining \eqref{eq:hierarchical_prior} with the likelihood gives the joint hierarchical posterior:
\begin{equation}
\label{eq:hierarchical_posterior}
p(\bx,\bx_{\mathrm{LR}}\mid\by) \propto p(\by\mid\bx)\,p_{\phi}(\bx \mid \bx_{\mathrm{LR}}) p_{\theta}(\bx_{\mathrm{LR}}).
\end{equation}
We extend the ancestral sampling from the prior setting to the posterior \eqref{eq:hierarchical_posterior}, where we first draw $\hat{\bx}_{\mathrm{LR}}$ from the marginal $p(\bx_{\mathrm{LR}}\mid\by)$, and then 
draw $\hat{\bx}$ from $p(\bx\mid\by,\hat{\bx}_{\mathrm{LR}})$.
Fig. \ref{posterior_learning} provides a pictorial representation of the proposed method.

\subsection{Stage 1: Low-resolution image reconstruction}
In the first stage, we recover a low-resolution image $\bx_{\mathrm{LR}}$ by sampling from the marginal of \eqref{eq:hierarchical_posterior}. We split the measurements into the central k-space samples $\by_{\mathrm{LR}}=\mathcal{P}\by$ and the remaining samples $\by_{\mathrm{HF}}$. Because the noise is independent, the marginal is
\begin{equation}
p(\bx_{\mathrm{LR}}\mid\by) \propto p(\by_{\mathrm{LR}}\mid\bx_{\mathrm{LR}})\, p_{\theta}(\bx_{\mathrm{LR}})\, Z(\bx_{\mathrm{LR}}),
\label{eq:lr_posterior}
\end{equation}
where $Z(\bx_{\mathrm{LR}})=\int p(\by_{\mathrm{HF}}\mid\bx)\,p_{\phi}(\bx\mid\bx_{\mathrm{LR}})\,d\bx$. 

In \eqref{eq:lr_posterior}, we approximate $p(\by_{\mathrm{LR}}\mid\bx)$ by $p(\by_{\mathrm{LR}}\mid\bx_{\mathrm{LR}})$ and treat $Z$ as constant, which are reasonable in accelerated MRI, where the periphery of k-space is sparsely sampled and therefore carries little information about $\bx_{\mathrm{LR}}$. 

The low-frequency k-space measurements and corresponding low-resolution images are related through the forward operator $\widetilde{\bA}=c\bS_{\mathrm{LR}}\bF_{\mathrm{LR}}\bC_{\mathrm{LR}}$ where $c=4$, $\bS_{\mathrm{LR}}$ is the central crop of the original acquisition mask, and $\bC_{\mathrm{LR}}$ is obtained by $c \times c $ average pooling of the coil sensitivity maps. The likelihood is then $p(\by_{\mathrm{LR}} \mid \bx_{\mathrm{LR}})\propto\exp\left(-\frac{1}{2\eta^2}\left\|\by_{\mathrm{LR}}-\widetilde{\bA}\bx_{\mathrm{LR}}\right\|_2^2\right)$. The first-stage reconstruction therefore produces a
measurement-informed low-resolution estimate
\begin{equation}
    \widehat{\bx}_{\mathrm{LR}}
    \sim
    p(\bx_{\mathrm{LR}} \mid \by_{\mathrm{LR}}).
    \label{eq:lr_reconstruction}
\end{equation}
We note that the low-resolution reconstruction is a reduced-dimensional inverse
problem with 16$\times$ fewer unknowns than the full-resolution
problem. The reduced dimensionality reduces the computational cost.
\subsection{Stage 2: High-resolution image reconstruction}
In the second stage, $\widehat{\bx}_{\mathrm{LR}}$ is used to condition a high-resolution diffusion prior. The final estimate is sampled from
\begin{equation}
    \widehat{\bx} \sim p(\bx \mid \by,\widehat{\bx}_{\mathrm{LR}}) \propto p(\by \mid \bx)\, p_{\phi}(\bx \mid \widehat{\bx}_{\mathrm{LR}}),
    \label{eq:hr_reconstruction}
\end{equation}
where $p(\by \mid \bx)$ enforces consistency with the original undersampled multi-coil k-space measurements and $p_{\phi}(\bx \mid \widehat{\bx}_{\mathrm{LR}})$ is the learned conditional high-resolution diffusion prior.
The low-resolution condition provides coarse anatomical and structural information, thereby restricting the set of plausible high-resolution images.
Consequently, the conditional prior models a more constrained
distribution than an unconditional high-resolution prior, which reduces the reconstruction ambiguity.  Although any sampling algorithm can be used, we sample from \eqref{eq:lr_reconstruction} and \eqref{eq:hr_reconstruction} using ALPS \cite{alps}. 
\section{Results}
\subsection{Dataset}
We evaluate the proposed cascaded diffusion framework on two-dimensional accelerated parallel T2-weighted MRI reconstruction. The dataset was obtained from the publicly available fastMRI brain dataset \cite{knee_dataset}. It
consists of sixteen-channel brain data with complex images
of size $384 \times 384$.  We evaluate the reconstruction performance across different acceleration factors and sampling geometries, including one-dimensional Cartesian and two-dimensional undersampling patterns.
\subsection{Architecture and Implementation}
\label{sec:impl}
\subsubsection{Cascaded Diffusion Model}
We trained a low-resolution diffusion prior on $96\times96$ images obtained by applying $4\times4$ average pooling with a stride of 4 to the original $384\times384$ training images. We also trained a high-resolution diffusion prior conditioned on the corresponding low-resolution images. Each conditioning image is interpolated to $384\times384$ and concatenated with the high-resolution image along the channel dimension. Following \cite{ho2022cascaded}, we apply conditioning augmentation to the low-resolution inputs during training to improve robustness to imperfect low-resolution conditions at inference. Because the conditioning images are obtained directly from the training data, the two models can be trained in parallel. Both priors use a SongUNet architecture without attention, with three resolution levels containing 64, 128, and 128 feature channels, respectively. Both models were trained using the EDM hyperparameters, loss function, and training noise-level distribution. The low-resolution unconditional network and the high-resolution conditional network, each has around 6.7 million parameters. 

\subsubsection{State-of-the-art (SOTA) methods for comparison}


We evaluate the performance of the cascaded diffusion model against the unconditional diffusion model-based algorithms. In particular, we compare with ALPS \cite{alps}, DAPS \cite{daps}, and DPS \cite{dps}. Each algorithm was implemented using the settings recommended in the respective paper.  We used a pre-trained unconditional diffusion model that follows the DDPM++ architecture. We call this model unconditional, although it takes the noise level as input. This model was trained using the EDM hyperparameters, loss function, and noise schedule. The number of model parameters is 65.5 million, which is around 10$\times$ the size of the cascaded networks and it operates at $384\times384$ image size.



\begin{figure}[t]
        \centering
        \includegraphics[width=0.8\linewidth]
        {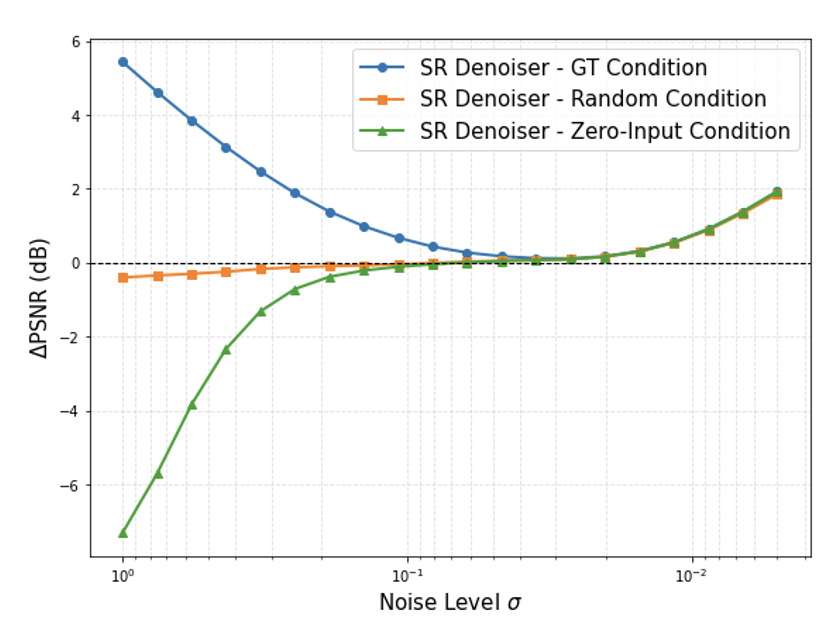}
    \vspace{-1em}
    \caption{\small Denoising performance of the the fine-scale conditioned diffusion model in cascaded model under different conditioning inputs, in comparison with the unconditional diffusion model. We observe that the ground truth condition (GT) results in a significantly improved performance at high noise scales compared to random or zero conditions. The coarse scale guidance offered by this model translates to improved performance }
    \label{fig:denoising_performance}
\end{figure}

\begin{table*}[t]
\centering
\caption{Reconstruction performance at different acceleration factors.}
\label{tab:reconstruction_results}
\setlength{\tabcolsep}{3pt}
\renewcommand{\arraystretch}{1}
\begin{tabular}{l|cc|cc|cc|cc}
\hline
{Acceleration} &
\multicolumn{2}{c|}{{Cascaded Model}} &
\multicolumn{2}{c|}{{ALPS: Diffusion}} &
\multicolumn{2}{c|}{{DPS: Diffusion}} &
\multicolumn{2}{c}{{DAPS: Diffusion}} \\
&
{PSNR (dB)} & {SSIM} &
{PSNR (dB)} & {SSIM} &
{PSNR (dB)} & {SSIM} &
{PSNR (dB)} & {SSIM} \\
\hline
1D $15\times$  & 28.10 $\pm$ 0.54 & 0.88 $\pm$ 0.02 & 23.06 $\pm$ 1.14 & 0.79$\pm$ 0.04 & 18.22 $\pm$ 0.97 & 0.51 $\pm$ 0.06 & 23.91$\pm$ 0.91 & 0.77 $\pm$ 0.02 \\
2D $15\times$  & 32.53$\pm$ 0.78  &0.95 $\pm$ 0.01 & 26.54 $\pm$ 1.07 & 0.89 $\pm$ 0.01 & 17.58 $\pm$ 0.58 & 0.48 $\pm$ 0.04 & 24.98$\pm$ 1.04 & 0.81 $\pm$ 0.02 \\
2D $10\times$ & 38.61 $\pm$0.75 & 0.96 $\pm$ 0.01 & 38.60 $\pm$ 0.67 & 0.96 $\pm$ 0.006 & 19.55 $\pm$ 0.73 & 0.58 $\pm$ 0.05 & 29.38 $\pm$ 0.74 & 0.86 $\pm$ 0.02 \\
\hline
\end{tabular}\vspace{1em}
\end{table*}
\subsection{Denoising Performance}



We evaluate the denoising performance of the cascaded conditional diffusion model using the downsampled reference image, random, and zero-valued low-resolution conditions. Figure~\ref{fig:denoising_performance} shows the PSNR difference relative to the unconditional diffusion model. At low noise levels, conditional and unconditional diffusion models perform similarly. As the noise level increases, informative conditioning provides an increasingly large improvement, whereas random or zero conditioning provides little benefit. This demonstrates that the advantage arises from informative low-resolution structure and becomes stronger as the denoising problem becomes more ill-posed.

\subsection{Reconstruction performance}

\begin{figure}[t]
    \centering
    \begin{subfigure}[t]{0.48\textwidth}
        \centering
        \includegraphics[width=0.87\linewidth]
        {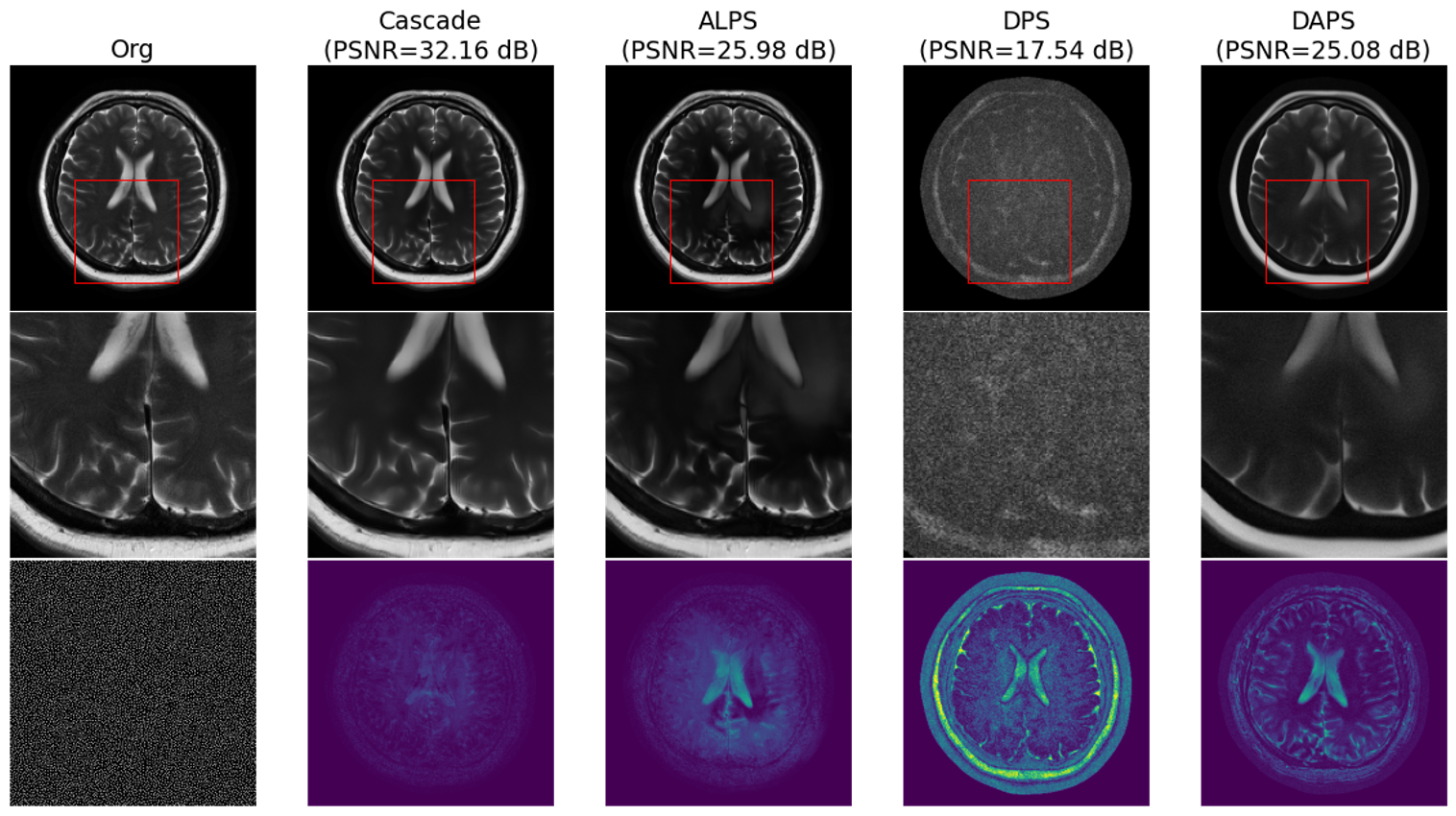}
        \caption{}
        \label{fig:acc_15x2D}
    \end{subfigure}
    \begin{subfigure}[t]{0.48\textwidth}
        \centering
        \includegraphics[width=0.87\linewidth]
        {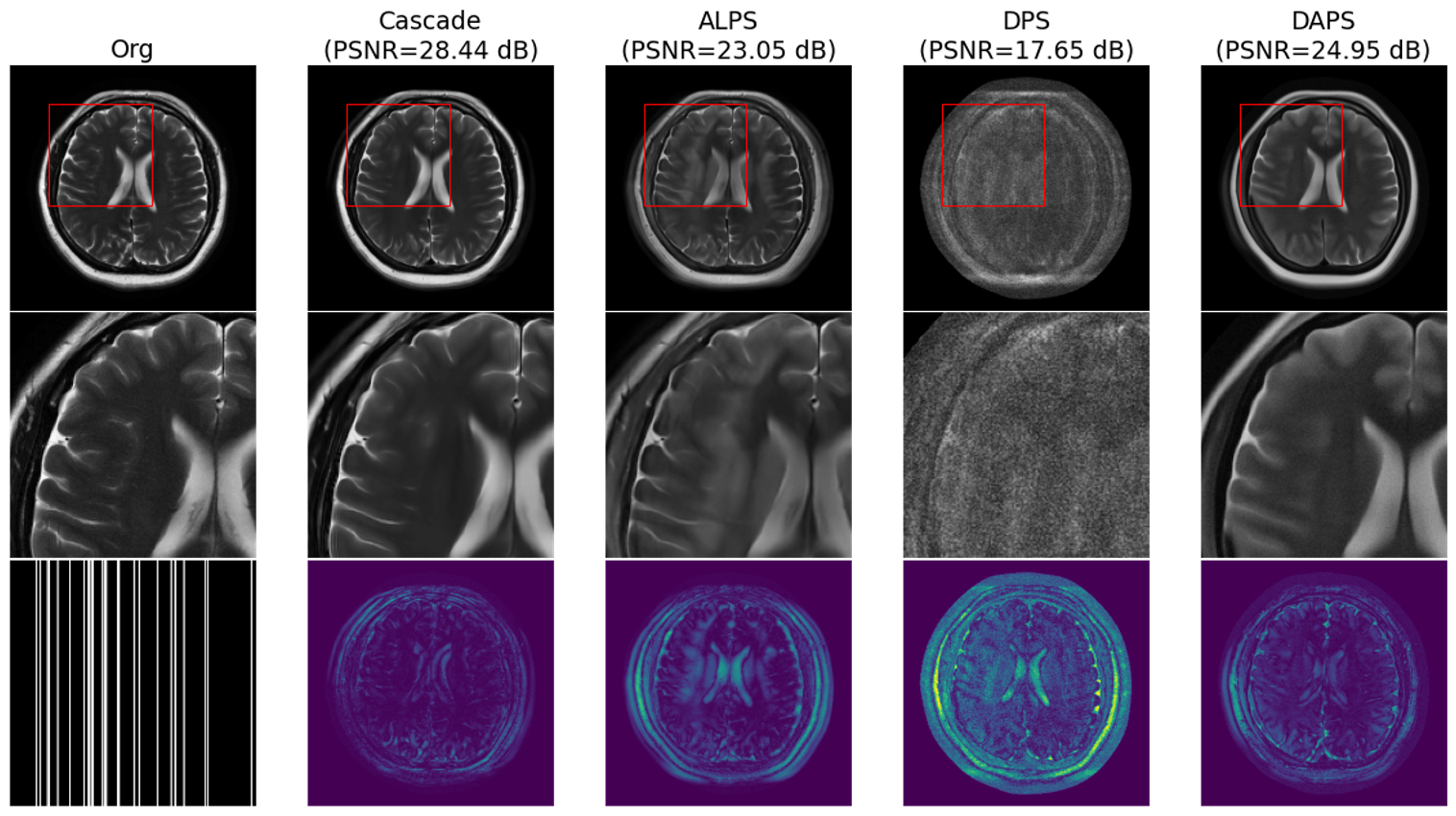}
       \caption{}
        \label{fig:acc_15x1D}
    \end{subfigure}

    \caption{Reconstruction performance of the cascaded diffusion model conditioned on the low-resolution image recovered by sampling from \eqref{eq:lr_reconstruction} and unconditional diffusion model for (a) 2D 15$\times$ and (b) 1D 15$\times$ acceleration. The top, middle, and bottom rows show the reconstructed images, magnified regions, and corresponding error maps, respectively.}
\label{fig:reconstruction_performance}
\end{figure}

\begin{figure}[t]
    \centering
    \begin{subfigure}[t]{0.4\textwidth}
        \centering
        \includegraphics[width=0.87\linewidth]
        {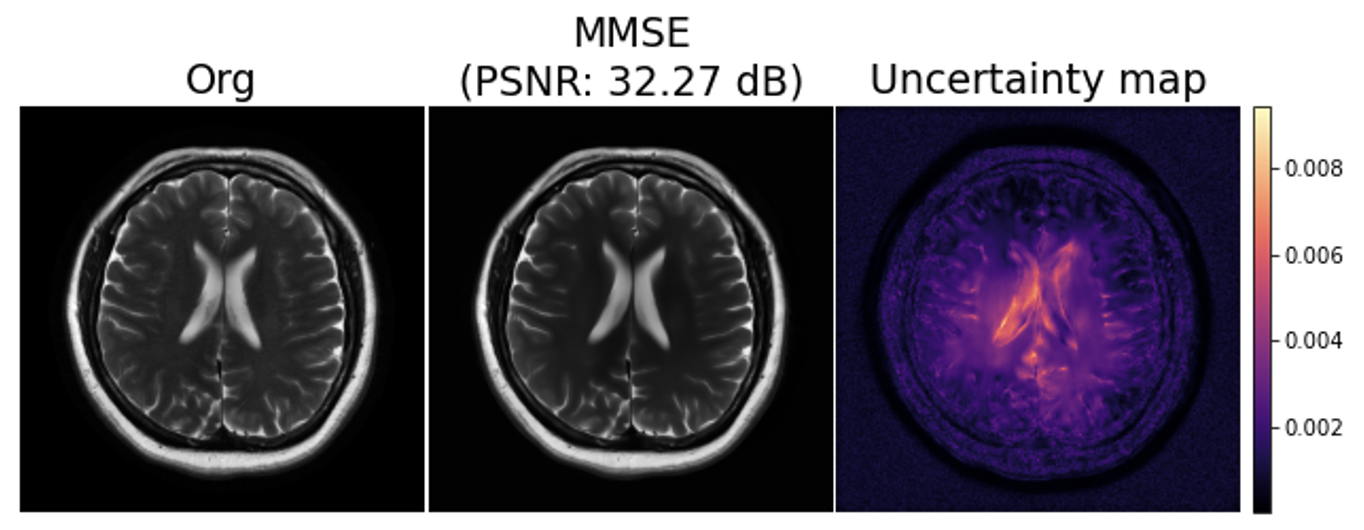}
        \caption{}
        \label{fig:acc_15x2D_sd}
    \end{subfigure}
    \hfill
    \begin{subfigure}[t]{0.4\textwidth}
        \centering
        \includegraphics[width=0.87\linewidth]
        {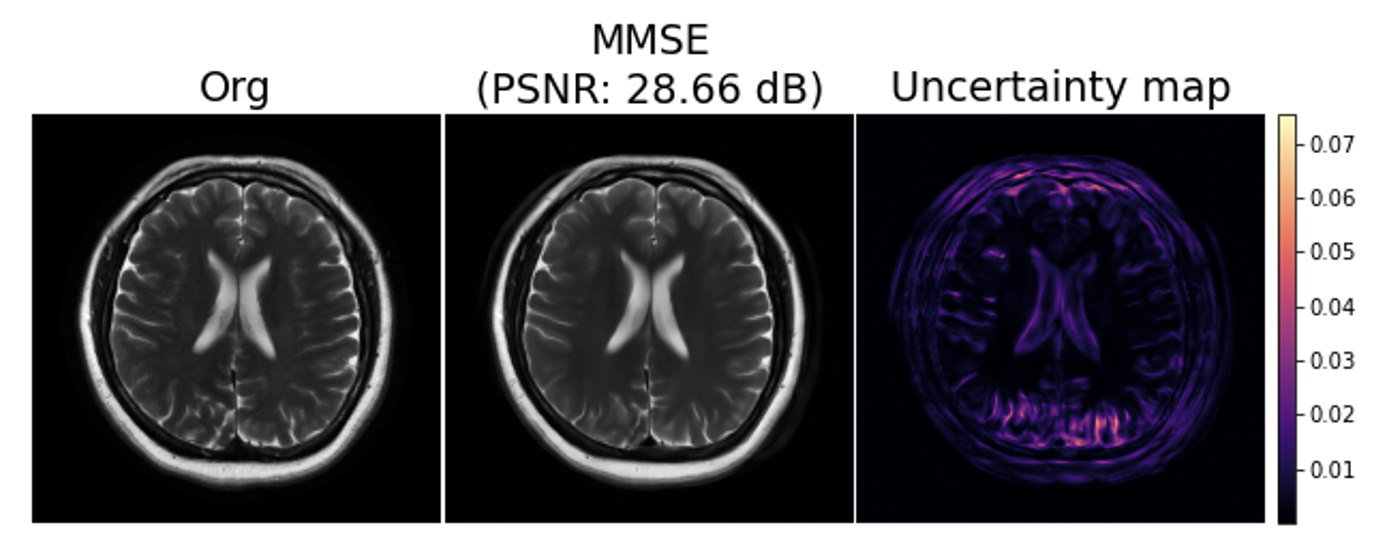}
       \caption{}
        \label{fig:acc_15x1D_sd}
    \end{subfigure}
    \caption{MMSE estimate and uncertainty map of the cascaded diffusion model for (a) 2D 15$\times$ and (b) 1D 15$\times$ acceleration. The uncertainty for the 2D 15$\times$ sampling pattern is approximately 10$\times$ lower than that for the 1D 15$\times$ sampling pattern.}
\label{fig:uncertainty}
\end{figure}

Table \ref{tab:reconstruction_results} compares the cascade with the unconditional-prior methods at 1D $15\times$, 2D $15\times$, and 2D $10\times$ acceleration. For each of the ten test slices, we average five posterior samples to estimate the posterior mean. At $15\times$ acceleration, the cascade outperforms the best baseline by 4.2~dB (1D) and 6.0~dB (2D) in PSNR, whereas at 2D $10\times$ it matches ALPS. This trend agrees with the denoising experiments, where informative conditioning helps most at low SNR. Fig.~\ref{fig:reconstruction_performance} shows reconstructions at $15\times$ acceleration, and Fig.~\ref{fig:uncertainty} shows the corresponding uncertainty maps, which are low across most of the image.


\section{Conclusion}
\label{sec:conclusion}


We introduced a cascaded posterior sampling framework that uses diffusion models as measurement-informed multiscale priors. The framework recovers a coarse image from the densely sampled center of k-space and uses it to condition a fine-scale prior. The coarse scale is thus well determined by the data, and the learned prior mainly resolves fine-scale detail, where the data are weak. In multi-coil MRI, the method improved on the best baseline by 4--6~dB in PSNR at $15\times$ acceleration, with networks that are each about $10\times$ smaller and faster to train. Because both priors are trained on images alone, they apply to any sampling pattern, and the data consistency remains linear at every scale. These properties make the approach attractive for 3D and high-resolution imaging, where full-resolution priors are costly to train and data are scarce. Future work will extend the cascade to more scales and to joint inference across scales.

\section{Compliance with ethical standards}

This study used a publicly available human subject dataset. Ethical approval was not required, as confirmed by its open-access license.

\noindent\textbf{Acknowledgments.} This work is supported by NIH grants R01-AG067078, R01-EB031169, and R01-EB019961.

\newpage
\bibliographystyle{IEEEbib}



\bibliography{strings,refs_new}

\end{document}